# Recuperation of Regenerative Braking Energy in Electric Rail Transit Systems

Mahdiyeh Khodaparastan, *Student Member, IEEE*, Ahmed A. Mohamed, *Senior Member, IEEE* and Werner Brandauer, *Member, IEEE*

*Abstract*—**Electric rail transit systems are large consumers of energy. In trains with regenerative braking capability, a fraction of the energy used to power a train is regenerated during braking. This regenerated energy, if not properly captured, is typically dumped in the form of heat to avoid overvoltage. Finding a way to recuperate regenerative braking energy can result in economic as well as technical merits. In this comprehensive paper, the various methods and technologies that were proposed for regenerative energy recuperation have been analyzed, investigated and compared. These technologies include: train timetable optimization, energy storage systems (onboard and wayside), and reversible substations.**

*Index Terms*— **Onboard energy storage, regenerative braking, reversible substation, wayside energy storage.**

## I. INTRODUCTION

Increasing the overall efficiency of electric rail transit systems is critical to achieve energy saving, and greenhouse gas (GHG) emission reduction [1], [2]. In general, electric train operation can be divided into four modes: acceleration, cruising, coasting and braking [3]. During the acceleration mode, a train accelerates and draws energy from a catenary or a third rail (i.e. a power supply rail, located next to the traction rails). In the cruising mode, the power of the motor is almost constant. In the coasting mode, the speed of the train is nearly constant, and it draws a negligible amount of power. In the braking mode, the train decelerates until it stops. In light rail traction systems and in urban areas, since the distance between the passenger stations is short, the cruising mode is typically omitted.

There are several types of train braking systems, including regenerative braking, resistance braking and air braking. In regenerative braking, which is common in today's electric rail systems, a train decelerates by reversing the operation of its motors. During braking, the motors of a train act as generators converting mechanical energy to electrical energy. In this paper, the produced electrical energy will be referred to as "regenerative braking energy" or "regenerative energy." This energy is used to supply train's onboard auxiliary loads, while the surplus energy is fed back to the third rail. In dense cities, the distance between passenger stations is typically short and train acceleration/braking cycles repeat frequently, which results in considerable amounts of regenerative energy [3].

Regenerative braking energy that is fed back to the third rail by a braking train can be utilized by neighboring trains that might be accelerating within the same power supply section as the braking one. However, this involves a high level of uncertainty since there is no guarantee that a train will be accelerating at the same time and location when/where regenerative energy is available. The amount of energy that can be reused by the neighboring trains depends on several factors, such as train headway and age of the system. If there are no nearby trains to use this regenerated energy, which is typically the case, the voltage of the third rail tends to increase. There is an over-voltage limit to protect equipment in the rail transit system. To adhere to this limit, a braking train may not be able to inject its regenerative energy to the third rail. The excess energy must be dissipated in the form of heat in onboard or wayside dumping resistors. This wasted heat warms up the tunnel and substation, and must be managed through a ventilation system [4].

Several solutions have been proposed in the literature to maximize the reuse of regenerative braking energy: (1) train timetable optimization, in which synchronization of multiple trains operation has been investigated. By synchronizing trains operation, when a train is braking and feeding regenerative energy back to the third rail, another train is simultaneously accelerating and absorbing this energy from the third rail; (2) energy storage systems (ESS), in which regenerative braking energy is stored in an electric storage medium, such as super capacitor, battery and flywheel, and released to the third rail when demanded. The storage medium can be placed on board the vehicle or beside the third rail, i.e. wayside; (3) reversible substation, in which a path is provided for regenerative energy to flow in reverse direction and feed power back to the main AC grid.

The goal of this paper is to provide a comprehensive review on the research efforts, studies and implementations that have been presented by both the academia and the industry on maximizing reuse of regenerative braking energy. Various solutions and technologies have been described and discussed. Advantages and disadvantages of each solution have been presented.

The rest of this paper is organized as follows. In section II, a discussion on system integration is presented, including the common topologies of rectifier substations. In section III, train timetable optimization is discussed. In section IV, the utilization of energy storage systems for regenerative energy recuperation in electric transit systems is discussed. In section

V, a brief guide to choosing the most suitable regenerative energy recuperation technique for a given transit system is presented. In section VI, a review on various methods and tools that have been used for modeling and simulating electric rail systems, is presented. Section VIII mainly focuses on nontechnical aspects related to the recuperation of regenerative braking energy. Finally, some of the conclusions that can be derived from this report are summarized in section IX.

## II. SYSTEM INTEGRATION

Electric rail transit systems consist of a network of rails, supplied by geographically distributed power supply substations. A typical DC transit substation consists of a voltage transformation stage that steps down medium voltage to a lower voltage level, followed by an AC/DC rectification stage that provides DC power to the third rail. There are also traction network protection devices (circuit breaker, insulator, etc.) both at the AC and DC sides to prevent personnel injuries and equipment damage. Transformers have overcurrent, Bochholtz and temperature protection and supply cables have differential feeder protection. On the DC side, rectifiers have overcurrent, reverse current trip protection and high speed breaker. There are impedance relays along the track or on the vehicle to protect earth faults. Figure 1 shows an example of a substation in which two transformers and two rectifiers are connected in parallel, to increase the power supply reliability. Auxiliary loads, such as elevators, escalator, ventilation systems and lighting systems are supplied through a separate transformer (referred to as "AUX TRANS" in Figure 1). A typical substation rating is 3 MW at 750 V DC supplying 4 kA with overload capabilities of 150, 300 and 4500 of the rated current for 1 hour, 1 minute and 10 s, respectively [5].

There are three common voltage levels for the third rail in DC transit systems: 600V, 750V and 1500V [6]. The operating third rail voltage is maintained between safety under/over voltage limits, e.g. ~500 and ~900 for a 750V system [7]. Based on IEEE Standard 1159-2009, voltage events in electric rail system can be classified as: 1) normal voltage that is considered 10% above and below the nominal value; 2) transient overvoltage that is voltage above 10% of the nominal value but for a very short time i.e. 0.5 cycles to 30 cycles; 3) interruption, which is voltage below 10% of nominal value (momentary: 0.5 cycle to 3 seconds, temporary: 3 seconds to 1 minute and sustained: over 1 minute); 4) voltage swell which is a voltage deviation above 110% of nominal (momentary: 30 cycle to 3 seconds, temporary: 3 seconds to 1 minute); 5) voltage sag, which is voltage deviation between 10% and 90% of the nominal voltage (momentary: 30 cycle to 3 seconds, temporary: 3 seconds to 1 minute); and 6) over/under voltage, which is voltage deviation above/ below 10%-20% of the nominal voltage lasting more than 1 minute. In some dense stations, an ESS may be used for voltage regulation [8].

The railway line circuit consists of traction rails and a power rail; in order to have better operation, protection and maintenance, they are divided into sections. Each section can be supplied by only one substation at one side or two substations at both ends. The first case is suitable for systems with short distance and few vehicles while the second case is suitable for systems with longer distances and many vehicles [9]. Trains move on traction rails while receiving their power from either a third rail or an overhead line. In addition to providing the friction force needed by the wheels to propel train vehicles, traction rails can also provide a return path for power. When a train moves on traction rails, the resistances between the train and its departure and arrival passenger stations change based on the distance [10].

In urban areas, the average distance between passenger stations is short (e.g. 1-2 mile). Therefore, trains accelerate rapidly to their maximum speed (e.g. 80-100 km/h), and decelerate shortly afterwards to prepare for their next stop. The typical average acceleration and deceleration rates are $1.1 \text{m/s}^2$ and $-1.3 \text{ m/s}^2$, respectively.

The electric vehicle is the main load of electric transit systems. The main electric part of a vehicle is its electric drive, which controls the torque and the speed of the vehicle. Electric drives are mostly consisting of converters and electric machines linked by appropriate control circuits. Specifically, there are DC-DC and DC-AC converters, DC motor and three-phase induction motor. In DC traction system, DC motors are controlled by a bank of resistors and a DC-DC converter. Three phase induction motors are controlled through DC-AC converters. DC-AC converters can be either voltage source inverter (VSI) or current source inverter (CSI). VSI does not need a fixed voltage and can handle +20% and -50% variation of traction line voltage. CSI needs a chopper converter to maintain DC link current constant [11], [12].

Some of the important factors that need to be considered in the selection of drives are as follows: line voltage regulation, quality of the power absorbed by the trains, electromagnetic interference introduced by the drive to the power line, preventing resonance that may be caused by the interaction between the drive and line components [11].

## III. TRAIN TIMETABLE OPTIMIZATION

Train timetable optimization has been proposed as one of the approaches to maximize the reuse of regenerative braking energy. In this method, the braking and acceleration actions of two neighboring trains are scheduled to occur simultaneously; therefore, some of the energy produced by the decelerating train is used by an accelerating one. Some studies show that up to 14% of energy saving can be achieved through timetable optimization [13]–[15].

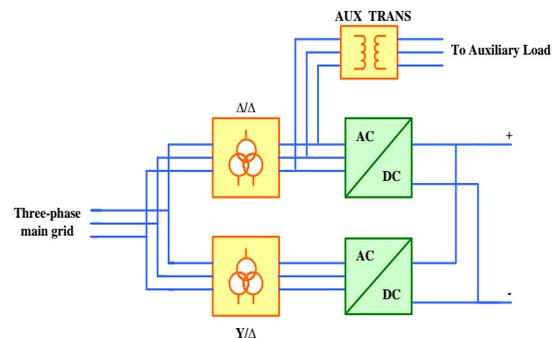

Fig. 1. A schematic diagram of a typical power supply substation

Studies that have been performed on train timetable optimization can be classified, according to their objectives, into two main categories: minimizing peak power demand, and maximizing the utilization of regenerative braking energy [16]. In the early stages of research on timetable optimization (i.e. early 1960s), the emphasis was on peak power demand reduction. Most of this research proposed methods to shift the acceleration time of some trains to off-peak time (note that time synchronization between trains was not targeted) [17], [18]. For instance, in [17], to limit the number of trains accelerating at the same time, train scheduling tables have been optimized using genetic algorithm. In [19], a control algorithm for coordinating movement of multiple trains has been proposed to reduce peak power demand. In [14], the peak power has been reduced through controlling train running time using dynamic programing.

More recently, research aimed at using train timetable optimization to maximize the utilization of regenerative braking energy, by synchronizing the acceleration/deceleration intervals of neighboring trains. Most of this research aimed at optimizing the dwell time (i.e. stop time at each station) of the trains to increase the chance of synchronizing accelerating and decelerating intervals [13], [15], [20]–[22]. Some other research focused on determining the optimal time overlap between multiple trains [23]–[25].

The optimization problem may be formulated to maximize utilization of regenerative braking energy, by finding the optimal departure and arrival times of trains. As an example, if a set of arrival times is considered as $a = \left\{ a_n^i, 1 \le i \le I, 1 \le n \le N \right\}$, where $a_n^i$ denotes the time that train $i$ arrives at station $n$, $I$ and $N$ are the number of trains and stations, respectively; and departure times are defined as $d = \left\{ d_n^i, 1 \le i \le I, 1 \le n \le N \right\}$, where $d_n^i$ denotes the time that train $i$ departs station $n$. The objective function of timetable optimization aimed to maximize utilization of regenerative braking energy can be formulated as follows:

$$F(a,d) = \sum_{t=1}^{T} \sum_{i=1}^{N_s} \min \left\{ \sum_{i=1}^{I} \omega_i(a,d,t) \lambda(i,t,s) \times \sum_{i=1}^{I} f_i(a,d,t) \lambda(i,t,s) \right\}$$

Where $T$ is the total operation time, $N_s$ is the number of electricity supply substations, $\omega_i(a,d,t)$ is the energy produced by train $i$ during the time unit $[t, t+1]$, $f_i(a,d,t)$ is the required energy for accelerating train $i$ during time unit $[t, t+1]$, and $\lambda(i,t,s)$ denotes whether the train $i$ is located in the electricity supply interval $s$ at time $t$ or not [16], [23].

Currently, there is ongoing research on integrated optimization methods, which combine train timetable optimization and speed profile optimization. Speed profile optimization is one of the conventional approaches used to improve the energy efficiency of electric rail transit system. In this approach, the speed profile of a single train is optimized such that it consumes less energy during the trips between stations. Timetable and speed profile optimization problems are tied to each other. Timetable optimization provides the best running time that can be used as an input in speed profile optimization. Simultaneously, speed profile optimization determines the optimal acceleration/deceleration rates, which can be used as an input to timetable optimization. An example of the integrated optimization method is presented in [26].

In this paper, the optimal dwell time at each station, and maximum train speed at each section is determined. The results show that 7.31% energy saving can be achieved using this approach.

The integrated optimization method provides better energy saving as compared to using timetable optimization or speed profile optimization individually. In [27], an integrated optimization method has been proposed based on actual operation data from Beijing Metro. The results show that the proposed method can reduce energy consumption of the overall system by 21.17% more than the timetable optimization method [28], and 6.35% more than the speed profile optimization method, for the same system and headways [16], [23]. In [29], other real world factors, such as the constant number of performing trains and cycle time are considered as part of the integrated optimization. For better utilization of regenerative braking energy, all trains supplied from the same electric section are considered in the time table. Another integration optimization method is presented in [30]. The aim of this paper is to optimize substations' energy consumption through finding optimal train movement mode sequences, inter-station journey times, and service intervals. An overview on some of the studies that is carried out in this area has been presented in Table I.

TABLE I
EXAMPLES OF TRAIN TIMETABLE OPTIMIZATION IMPLEMENTATIONS

| Method | Saving (% of consumed energy) | Implemented in real system? | Comment | Ref. |
|---|---|---|---|---|
| Dwell time optimization with GA algorithm | 14% | No, but data are gathered from Tehran Metro | The impact of both headway and dwell time on reusing regenerative energy has been studied. | [13] |
| Running time reserve optimization with GA | 4% | No, but data are gathered from Berlin Metro | The headway is considered to be constant. At each stop, the amount of reserve time to be spent on the next section of the ride is decided. | [14] |
| Dwell time optimization by greedy heuristic method | 5.1% | No | A mathematical model of metro timetable has been defined. | [15] |
| Departure time through multi-criteria mixed integer programing | - | No, but data are gathered from a Korean subway system | Around 40% of peak energy has been reduced and utilization of regenerative braking energy is improved by 5%. | [20] |
| Departure and dwell time Optimization | 7% in simulation, 3.52% in reality | Proposed models were used to design a timetable for Madrid underground system | 85% of braking and accelerating processes are synchronized. | [21] |
| Fuzzy logic control Dwell time | Not presented. | No | Train operation is specified by a set of indices, and the aim of fuzzy control is to find the best performance among these indices. | [22] |
| Running time Optimization by GA | 7% | No, but data are gathered from a substation in the UK | Two objective functions are considered: energy consumption and journey time. The best possible compromise between them is searched using GA. | [114] |
| Direct climbing optimization | 14% | No, but data are gathered from a substation in Italy | The optimal set of speed profile and timetable variables has been found in order to minimize energy consumption. | [115] |
| Dwell time and run time optimization/Genetic algorithm and dynamic programing model | 6.6% | - | From energy saving point of view, run-time control is superior over dwell time control. More flexible train control can also be achieved with run time control. | [91] |
| Substation energy consumption optimization | 38.6% | No, but the case study was based on Beijing Yizhuang Metro | Substation energy consumption optimized by modifying the speed profile and the dwell time | [116] |

## IV. STORAGE BASED SOLUTIONS

An energy storage system, if properly designed, can capture the energy produced by a braking train and discharge it when needed. Consequently, the amount of energy consumed from the main grid is reduced [20], [21], [31].

In addition, using ESS can reduce the peak power demand, which not only benefits the rail transit system but also the power utility. ESS may be used to provide services to the main grid, such as peak shaving [4].

Since the energy regenerated by a braking train is captured by an ESS, the need for onboard or wayside dumping resistors is minimized. Therefore, heat waste and ventilation system costs are reduced [22].

ESS can be implemented in two different ways: onboard and wayside. In onboard, ESS is mostly located on the roof of each train. On the other hand, wayside ESS is located outside the train, on the trackside. It can absorb the regenerative energy produced by all trains braking within the same section and deliver it later to other trains accelerating nearby. Both wayside and onboard ESS will be described in the following subsections, and examples of their application in transit systems all over the world will be presented. Before that, the common technologies available and used for ESS in the rail transit system will be briefly discussed.

### A. Energy storage Technologies

Selection of the most suitable storage technology is a key factor to achieve optimal ESS performance for a given application. Several important factors must be considered while designing an ESS, and choosing the most suitable storage technology. These factors include: the energy capacity and specific energy, rate of charge and discharge, durability and life cycle [7]. The common energy storage technologies that have been utilized in rail transit systems are batteries, supercapacitors and flywheels.

#### 1) Batteries

Battery is the oldest electric energy storage technology, which is widely used in different applications. A battery consists of multiple electrochemical cells, connected in parallel and series to form a unit. Cells consist of two electrodes (i.e. anode and cathode) immersed in an electrolyte solution. Batteries work based on the following principle: due to reversible chemical reactions (i.e. oxidation and reduction) that occur at the electrodes, a potential difference appears between them (voltage between the anode and the cathode). Consequently, energy can reversibly change from the electrical form to the chemical form [32], [33].

There are various types of batteries depending on the material of their electrodes and electrolyte. Among those types, the most commonly used in rail transit systems are: Lead–acid (pbso$_4$), Lithium-ion (Li-ion), Nickel-metal hydride (Ni-MH) and sodium sulfur (Na-s). Other types of batteries like flow battery may have the potential to be used in rail transit systems [34][35]. A comparison of the advantages and disadvantages of each type has been summarized in Table II.

TABLE II
A COMPARISON OF DIFFERENT BATTERY TECHNOLOGIES

| Type | Advantages | Disadvantages | Comment | Reference |
|------|-----------|---------------|---------|-----------|
| Pbso4 | • Low cost per Wh<br>• Long history<br>• Wide deployment<br>• High reliability<br>• High power density | • Low number of cycle<br>• Low charging current<br>• Limited service life<br>• Environmental concern<br>• Poor performance in low temperature | -Recently, extensive research has been carried out on replacing lead with other materials, such as carbon, to increase its power and energy density | [32], [36], [74] |
| Ni-MH | • Long service life<br>• High energy<br>• High charge/discharge current<br>• High cycle durability<br>• Low Environmental concern | • High cost per Wh<br>• High maintenance<br>• High self-discharge rate. | -The main disadvantage is high self-discharge rate, might be overcome using novel separators | [32], [74], [117], [118] |
| Li-ion | • High energy density<br>• Being small and light<br>• Low maintenance<br>• High number of cycle | • High cost per Wh<br>• Require cell balance and control to avoid overcharge<br>• Required special packing and protection circuit | - Currently, researchers investigate a combination of electrochemical and nanostructures that can improve the performance of Li-ion batteries | [32], [74], [119] |
| Na-s | • High energy density<br>• High power density<br>• Highly Energy efficiency | • High cost<br>• Environmental concern<br>• Need cooling unit | -Researchers are investigating new ways to reduce their high operating temperature. | [32], [36], [117], [118], [120] |

*2) Flywheels*

Flywheel is an electromechanical ESS that stores and delivers kinetic energy when it is needed. Flywheel is composed of an electrical machine driving a rotating mass, so called rotor, spinning at a high speed. The amount of energy that can be stored or delivered depends on the inertia and speed of the rotating mass. During the charging process, the electrical machine acts as a motor and speeds up the rotor increasing the kinetic energy of the flywheel system. During the discharging process, the rotational speed of the rotor decreases releasing its stored energy through the electrical machine, which acts as a generator. The electrical machine is coupled to a variable frequency power converter. To reduce friction losses, flywheels use magnetic bearing, and to reduce air friction losses, the rotor is contained in a vacuum chamber [32], [33], [36], [37], [38].

Some of the advantages of flywheel ESS are high energy efficiency (~95%), high power density (5000 W/kg) and high energy density (>50 Wh/kg), less maintenance, high cycling capacity (more than 20000 cycles) and low environmental concerns [39]. Flywheel systems present some drawbacks, such as very high self-discharge current, risk of explosion in case of failure, high weight and cost. However, system safety is believed to be improvable through predictive designs, and smart protection schemes. According to some publications, if/when the cost of flywheel systems is lowered; they can be extensively used in all industries and play a significant role in the worldwide energy sustainability plans [33], [36], [40].
Based on the simulation results presented in [41], flywheel ESS is capable of achieving 31% energy saving in light rail transit systems.

*3) Super Capacitors*

Super capacitor is a type of electrochemical capacitors consisting of two porous electrodes immersed in an electrolyte solution. By applying voltage across the two electrodes, the electrolyte solution is polarized. Consequently, two thin layers of capacitive storage are created near each electrode. There is no chemical reaction, and the energy is stored electrostatically. Because of the porous electrode structure, the overall surface area of the electrode is considerably large. Therefore, the capacitance per unit volume of this type of capacitor is greater than the conventional capacitors [32], [36], [42]–[46].

The electrical characteristics of super capacitors highly depend on the selection of the electrolyte and electrode materials [43]. Super capacitors have several advantages, such as high energy efficiency (~95%), large charge/discharge current capacity, long lifecycle (>50000), high power density (>4000) and low heating losses [36], [43], [45], [39], [47]. However the maximum operating voltage of ultra-capacitors is very low and they suffer from high leakage current. Because of these two drawbacks, they cannot hold energy for a long time [42]. Recently, Li-ion capacitors have been developed with less leakage current and higher energy and power densities than batteries and standard super capacitors [42], [48], [49].

## B. Onboard Energy Storage

In onboard ESS, the storage medium is placed on the vehicle. It can be placed on the roof or under the floor of the vehicle. Placing ESS under the floor is relatively costly, because space is not readily available. The efficiency of onboard ESS is highly dependent on the characteristic of the vehicle, which can directly affect the amount of energy produced and consumed during braking and acceleration, respectively [50]. Other advantages of onboard energy storage are peak power reduction, voltage stabilization, catenary free operation and loss reduction. On the other hand, the cost of implementation, maintenance, and safety concerns, are high because unlike wayside storage, in onboard ESS, an ESS is needed for each train.

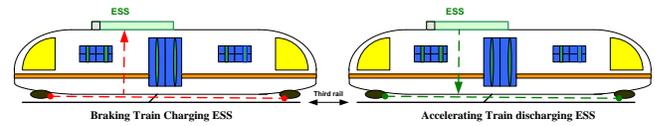

Fig. 2. Onboard energy storage systems.

Onboard ESS is already in use by some rail transit agencies. In addition, several agencies all around the world are considering –or actually testing- it. Various technologies have been used for onboard ESS; among them, super capacitors have been more widely implemented in many transit systems. Due to safety and cost limitations, onboard flywheels did not acquire much attention, and still need more investigation. However, there are some ongoing efforts. For instance, construction of a prototype for hybrid electric vehicle by CCM has been reported in [51]. An agreement between Alstom Transport and Williams Group on installation of onboard flywheel on trams has been reported in [52]. On the other hand, batteries have not been able to compete with super capacitors due to their short lifetime, and low power density.

Important examples of real world implementation of onboard ESS are Brussel metro and tram lines and Madrid Metro line in Europe that show 18.6%-35.8% and 24% energy saving, respectively [53], [54] [55]. Japan metro with 8% saving of regenerative braking energy, and Mannheim tramway with 19.4%-25.6% increase in the overall system energy efficiency are two other examples of real world implementation of onboard ESS [56], [57].

In academic research, studies mostly focus on optimal design, sizing and control of onboard ESS. For instance, in [58], an onboard super capacitor ESS control strategy integrated with motor drive control has been presented. A control method for maximum energy recovery has been presented in [59]. In this method, a line in Rome metro has been considered as a case study. Theoretical Results show 38% energy recovery. Table III provides an overview of various examples for onboard ESS worldwide.

## C. Wayside Energy Storage

A schematic overview of wayside ESS is shown in Fig. 3. The main concept of wayside ESS is to temporarily absorb the energy regenerated during train braking and deliver it back to the third rail when needed. Generally, it consists of a storage

TABLE III
EXAMPLES OF ONBOARD ENERGY STORAGE IMPLEMENTATIONS

| Type | Location | Purpose | Comment | Reference |
|------|----------|---------|---------|-----------|
| Ni-MH | Sapporo | Energy saving<br>Catenary free operation | Giga-cell NiMH batteries provided by Kawasaki has been used. It can be fully charged in five minutes through the 600V DC overhead catenary. | [70] |
| Li-ion | Charlotte | Energy saving<br>Catenary free operation. | -- | [70], [121], [122] |
| Ni-MH | Lisbon | Operation without overhead contact line | The SITRAS HES (hybrid energy storage) energy storage system has been used. | [6], [123] |
| Ni-MH | Nice | Catenary free operation. | - | [54], [91] |
| Super capacitor | Mannheim | Reduction of energy consumption and peak power demand.<br>Catenary free operation. | A 400V system with 1 kWh energy<br>640 Ultra-caps, with a capacity of 1800F each. | [64],[124]–[126] |
| Super capacitor | Innsbruck | Energy saving | - | [6] |
| Super capacitor | Seville, Saragossa | Energy saving, Catenary free operation. | - | [122] |
| Super capacitor | Paris | Energy saving, Catenary free operation | Could also be recharged from the overhead contact system in about 20 seconds during station stops. | [122], [127] |
| Flywheel | Rotterdam (France) | Energy saving<br>Catenary free operation | Flywheel located at the roof. Flywheel system was developed and installed by ALSTOM. However, the project stopped due to technical issue. | [70], [128] |
| Battery | Brookville | Catenary free operation. | - | [122] |

medium connected to the third rail through a power control unit [62].

In addition to the general advantages that were previously mentioned for energy storage systems, wayside ESS can also help minimize problems related to voltage sag [4], [50], [63]. Voltage sag, which is temporary voltage reduction below a certain limit for a short period of time, can damage electronic equipment in a rail car, and affect the performance of trains during acceleration. ESS can be designed to discharge very fast, and by injecting power to the third rail, they help regulate its voltage level [64]. In addition to the economic benefits provided by ESS through recapturing braking energy, ESS can be designed to participate in the local electricity markets as a distributed energy resource [65]. Some other applications that can be provided by wayside ESS include peak shaving, load shifting, emergency backup and frequency regulation [8].

In Madrid, an operating prototype is demonstrating the use of the rail system infrastructure including wayside ESS for charging electric vehicles [66].

Real world implementation of wayside ESS has reported energy savings of up to 30%. The amount of energy saving by ESS highly depends on the system characteristics and storage technology. As an example, the commercially available wayside ESS, Sitras SES (Static Energy Storage) system marketed by Siemens is presented as a solution that can save nearly 30% of energy. The proposed ESS use a supercapacitor technology that can provide 1MW peak power, and is capable of discharging 1400 A DC current into the third rail during 20-30 second. Sitras ESS is implemented in different cities in Germany (Dresden, Cologne, Koln and Bochum), Spain (Madrid) and China (Beijing). Bombardier has developed a system based on super capacitors, the EnerGstor, which is capable of offering 20% to 30% reduction in grid power consumption. An Energstor prototype, sized 1 kWh per unit, has been designed, assembled and tested at Kingston (Ontario) [6].

Another Supercapacitor-based system that is commercially available is Capapost, developed by Meiden and marketed by Envitech Energy, a member of the ABB Group, with scalability from 2.8 to 45 MJ of storable energy. This system has been reported to be installed in Hong Kong and Warsaw metro systems [67].

Table IV provides an overview of various applications of wayside ESS all over the world. This information is mostly published by manufacturers of wayside ESS like Siemens [6], ABB [65], VYCON [68], [69], Pillar [70].

### D. Energy Storage Control and Energy Management

The control and energy management of ESS play a critical role in rail system applications, due to the stringent time requirements (i.e., the frequent fast-charging/discharging cycles). The specific targeted service, e.g., energy saving or voltage regulation, changes how the ESS should optimally be controlled. Besides the research and development efforts that have been performed on the design and analysis of ESS, both onboard and wayside, some other research studies targeted optimal sizing and siting [60], [61][71], [72]; and ESS energy management and control. Some of the most significant efforts that were carried out in this field have been reviewed in Table V.

## V. REVERSIBLE SUBSTATION

Another approach to reuse regenerative braking energy is through the use of reversible substations, as shown in Fig. 4. A reversible substation, also known as bidirectional or inverting substation, provides a path through an inverter for regenerative braking energy to feed back to the upstream AC grid, to be consumed by other electric AC equipment in the substation, such as escalators, lighting systems, etc. [73]. This energy can also feed back to the main grid based on the legislations and rules of the electricity distribution network.

Reversible substations must maintain an acceptable power quality level for the power fed back to the grid by minimizing the harmonics level [73].

Even though reversible substations are designed to have the ability to feed regenerative braking energy back to the upstream network, if maximum regenerative energy recuperation is targeted, priority should be given to the energy exchange between trains on the DC side of the power network.

TABLE IV
EXAMPLES OF WAYSIDE ENERGY STORAGE IMPLEMENTATIONS

| Type | Location | Voltage | Purpose | Comment | Reference |
|---|---|---|---|---|---|
| Li-ion | Philadelphia | 660V | - Energy saving<br>- Optimize SEPTA's power and voltage quality<br>- Frequency Regulation Market Revenues | ENVILINE™ ESS provided by ABB has been used. | [65],[112] |
| NAS | Long island | 6kV AC | - Peak shaving | There were several challenges reported with this project, such as the sizing of ESS, safety issues, and unexpected costs. | [36],[117], [118], [120] |
| Li-ion | West Japan | 640V | - Energy saving<br>- Voltage stabilization | - | [129] |
| Li-ion | Nagoya | 640V | - Voltage stabilization | - | [129] |
| Li-ion | Kagoshima | 640V | - Voltage enhancement | The ESS was far from the substation and controlled remotely via internet. | [129] |
| Ni-Mh | Osaka | 640V | - Energy saving | The battery was connected directly to the electric line. | [129] |
| Li-ion | Kobe | 640V | - Voltage enhancement | - | [129] |
| Super Capacitor | Seibu | 640V | - Energy saving | - | [129] |
| Ni-MH | New York | 670V | - Voltage enhancement | The battery was directly connected to the third rail. | [130] |
| Flywheel | London | 630V | - Energy saving<br>- Voltage enhancement | Flywheel system provided by URENCO. | [36], [131], [132] |
| Flywheel | Los Angeles | - | - Energy saving | Flywheel system provided by VYCON | [68], [69] |
| Flywheel | Hanover | - | - Energy saving | Flywheel system provided by Pillar | [70] |
| Flywheel | New York | 670V | - Energy saving | Flywheel system was provided by KINETIC TRACTION, and successfully tested at Far Rockaway. However, the project was stopped due to budget constraints. | [70] |
| Super Capacitor | Madrid | 750V | - Voltage stabilization | Sitras SES has been used. | [6] |
| Super Capacitor | Cologne | 750V | - Energy saving, voltage stabilization | Sitras SES has been used | [6] |
| Super Capacitor | Beijing | 750V | - Energy saving | Sitras SES has been used | [6] |
| Super Capacitor | Toronto | 600V | - Energy saving | Sitras SES has been used | [6] |

There are two common ways to provide a reverse path for the energy: 1) using a DC/AC converter in combination with a diode rectifier; and 2) using a reversible thyristor-controlled rectifier (RTCR). In the first approach, the DC/AC converter can be either a pulse width modulation (PWM) converter, or thyristor line commutated inverter (TCI) [74]. It is worth mentioning that in the first approach, the existing diode rectifier and transformer can be kept, and some additional equipment needs to be added for reversible energy conduction. However, in the second approach, the diode rectifiers need to be replaced with RTCRs and the rectifier transformers need to be changed, which makes this approach more expensive and complex [74]. However, RTCRs have advantages, such as voltage regulation and fault current limitation [75].

A TCI is an anti-parallel thyristor controlled rectifier (TCR) connected backward to provide a path for transferring energy from the DC side to the AC one. This technology has been used in an Alstom reversible substation setup called HESOP (Harmonic and Energy Saving Optimizer) [76], as shown in Fig. 5. The rated current of the TCI is half of that of the forward TCR, which reduces its cost. To use a TCI with an existing circuit, an auto transformer and a DC reactor should be used to increase the AC voltage, and limit circulating currents between the TCI and the diode rectifier. To minimize AC harmonics, a 12 pulse system has been proposed in [74].

As mentioned above, a diode rectifier can also be combined with a PWM converter to provide a reverse path for the energy. PWM converters have the advantage of working at unity power factor, and the disadvantage of high cost and high switching losses. In order to use PWM converters for reversible substation purposes, a step up DC/DC converter should be added between the PWM converter and the DC bus.

TABLE V
CONTROL AND ENERGY MANAGEMENT

| Technology | Goal | Reference |
|---|---|---|
| Onboard SC | Line voltage control | [133] |
| Onboard SC | Maximum energy recovery | [134] |
| Onboard SC | Maximum energy recovery | [135] |
| Onboard SC | Maximum energy recovery | [61] |
| On board SC | Maximum energy recovery | [136] |
| Wayside SC | Power flow control | [137] |
| Wayside Li-C | Power flow control | [138] |
| Wayside SC | Maximum energy recovery | [139] |
| Wayside SC | Maximum energy recovery | [140] |
| Wayside SC | Maximum energy recovery | [141] |
| Wayside SC | Improve dynamic performance of system | [142] |
| Wayside SC | constant voltage-based energy management strategy | [143] |
| Wayside SC | Energy management control | [144] |
| Wayside Li-C | Energy management control | [108] |
| Wayside Li-C | Line voltage control | [145] |
| Flywheel | Line voltage control | [146] |
| Flywheel | Line voltage control | [147] |
| Battery | Peak load shifting | [148] |
| Battery | Maximum energy recovery | [149] |
| Hybrid bat-SC | Energy management control | [150] |

TABLE VI
COMPARISON OF DIFFERENT TYPES OF REVERSIBLE SUBSTATIONS

| Parameters | TCI$_{Sitras}$ | TCI$_{Ingeteam}$ | RTCR |
|---|---|---|---|
| P$_{regen\,peak}$ (MW) | 3 | 1.5 | 3 |
| P$_{rectifier\,peak}$ (MW) | 0 | 0 | 12 |
| Efficiency (%) | 96 | 92–94 | 97 |
| Size (m³) | $2.4 \times 1 \times 2.3$ | NA | $3.5 \times 0.8 \times 2.5$ |
| Footprint (m²) | 2.4 | 7.5 | 5.25 |
| Weight (t) | 2.65 | NA | NA |
| life (years) | 20+ | 20+ | 20+ |

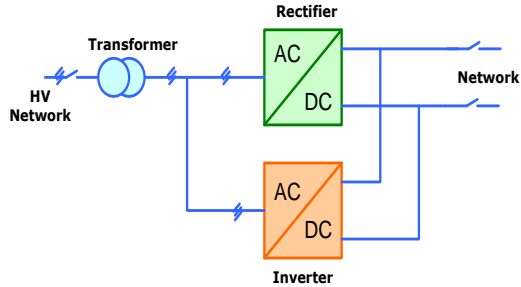

Fig. 4. Block diagram of a reversible substation.

In addition, to reduce the harmonics level and avoid current circulation, a DC filter needs to be added at the output of the converter.

A similar technology has been developed by INGEBER, where, an Inverter and a DC chopper (DC/DC converter) are connected in series, and their combination is connected to the existing substation [77], [78]. Fig. 6 shows a typical RTCR. It consists of two TCRs, which are connected in parallel, providing a path for the energy in forward and reverse directions. Only one of these TCRs can be fired at a time; therefore, no current will circulate between them, and there will be no need for a DC inductor. When RTCR is working in the forward direction (AC/DC), the inverter acts as an active filter.

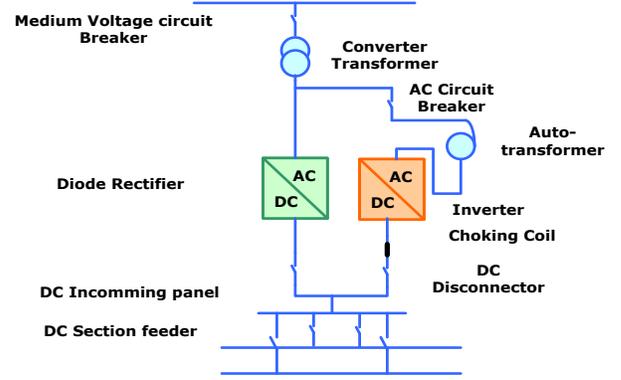

Fig. 5. Block diagram for HESOP reversible substation.

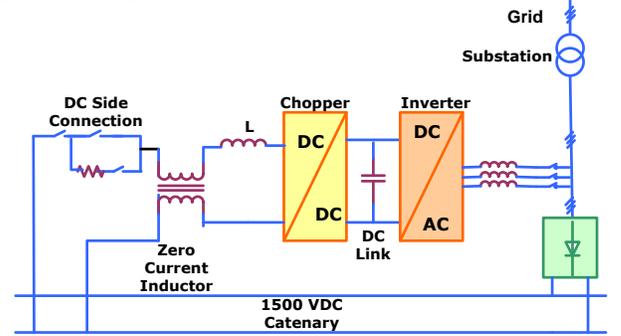

Fig. 6. Block diagram for diode rectifier/PWM converter reversible substation.

However, the rectifier only works in the traction mode [76]. In order to switch between the rectifier and inverter modes, a single controller provides pulses for both of them without any dead time [76] A comparison between the aforementioned reversible substation technologies is presented in Table VI [74]. Beside these technologies, ABB also proposed two technologies called Enviline TCR and Enviline ERS. Enviline TCR is a Traction Control Rectifier that uses four quadrant converters to provide a reverse path for energy flow in the substation. This technology can be connected in parallel with an existing diode rectifier in a substation [79]. Enviline ERS is a wayside energy recuperation system consisting of an IGBT based inverter that can be connected in parallel with the existing substation's rectifier to return surplus energy to the main grid, and can also be configured to work as a rectifier to boost rectification and provide reactive power support, if needed [80]. Some of the currently available reversible substation systems are summarized in Table VII.

## VI. CHOOSING THE RIGHT APPLICATION

Different technologies/techniques have been proposed to reuse regenerated braking energy. A comparison between the general pros and cons associated with these alternatives is presented in Table VIII [70], [74].

Choosing the right regenerative energy recuperation technique/technology requires careful consideration of various influential parameters, such as [39]:

- Catenary-free operation
- Electric network ownership
- Electric network characteristic

TABLE VII
EXAMPLES OF REVERSIBLE SUBSTATION IMPLEMENTATIONS

| Company | Location | Voltage Level | Technology | Energy Saving | Comment | Ref. |
|---------|----------|---------------|------------|---------------|---------|------|
| Alstom-HESOP | Paris Tramway | 750 | Thyristor rectifier bridge associated with an IGBT converter | 7% of traction consumed energy | - Recuperated more than 99% of recoverable braking energy. <br> - Reduced train heat dissipation, which led to reducing energy consumption used for ventilation. <br> - No need for onboard braking resistors; therefore, train mass was reduced leading to less energy consumption during acceleration. | [76], [151], [152] |
| | Utrecht–Zwolle | | | | | |
| | London | | | | | |
| | Milan Metro | 1500 | | | | |
| Simense-Sitras TCI | Oslo, Singapore | 750 | Inverter, B6 thyristor bridge, autotransformer in parallel with a diode rectifier | - | -Remote control is possible through a communication interface. <br> - Reduced the number of braking resistors on the train. <br> - In the 750V version, an autotransformer is integrated with the inverter, but in the 1500V version or high power 750V, an autotransformer is installed separately. | [70] |
| | Zugspitze, Germany | 1500 | | | | |
| Ingeber-Ingteam | Bilbao -Spain | 1500 | Inverter in parallel with existing rectifier and transformer | 13% of the substation annual energy consumption | - The current injected to the AC grid is high quality (THD < 3%). <br> - A DC/DC converter and an inverter are connected in series, and their combination is connected to an existing grid. <br> - Reduced carbon emissions by 40%. | [70], [77], [78], [153] |
| | Malaga -Spain | 3000 | | | | |
| | Bielefeld-Germany | 750 | | | | |
| | Metro Brussels. | 750 | | | | |

TABLE VIII
A COMPARISON BETWEEN DIFFERENT RECUPERATION TECHNIQUES

| Application | Advantages | Disadvantages |
|------------|-----------|---------------|
| Onboard ESS | -Provides possibility for catenary-free operation. <br> -Reduces voltage drop. <br> -Reduces third rail losses and increases efficiency. | -High cost due to placement of ESS on the vehicle. <br> -High safety constraints due to onboard passengers. <br> -Standstill vehicles for maintenance and repair. |
| Wayside ESS | -Mitigates voltage sag. <br> -Can be used by all vehicles running on the line (within the same section). <br> -Maintenance and repair do not impact train operation. | -Increases overhead line losses due to the absorption and release of energy over the traction line. <br> -Analysis is needed to choose the right sizing and location. |
| Reversible Substation | -Provides possibility for selling electricity to the main grid. <br> -Can be used by all vehicles running on the line. <br> -Maintenance and repair do not impact train operation. <br> -Lower safety constraints. | -No voltage stabilization. <br> -Analysis is needed for choosing the right location. |

- Vehicles
- Headways

Catenary-free operation provides an opportunity for vehicles, especially tramway, to run without connection to the overhead line. It is a good solution for operating trams in, for instance, historic areas. In this case, only onboard energy storage can be used, and can be charged through regenerative braking and/or fast charging infrastructure in the stations [39].When considering different applications of energy recovery, electric network ownership plays an important role.

In many cases, local energy providers own the power system infrastructure, and transportation companies have limited autonomy on the electric network. In this case, if there is a possibility to sell the recovered energy for a high price, reversible substations may be the best option. Otherwise, wayside energy storage should be selected [39].

The characteristics of the network represent another important parameter that should be considered when selecting the energy recovery technique. For example, in some stations with a high level of traffic, the energy exchange between braking and accelerating trains occurs naturally. Therefore, the energy recovery application should be designed in a way that gives priority to natural energy exchange between the trains.

Moreover, some stations face serious voltage drop (due to aging, etc.) when several trains accelerate simultaneously. In this case, wayside energy storage may be used to help sustain the third rail voltage [39].

The type of vehicle used in a transportation system is also an important factor. For example, in some systems, old and new vehicles may be running together, while old vehicle may not have the ability to regenerate energy during braking. Therefore, investing in energy recovery may not yield the same value. In addition, the weight of the vehicles affects the regenerative braking energy, such that heavier vehicle produces more energy during braking. An average rate of train occupancy should be considered during designing and analyzing the ESS system [39].

## VII. ELECTRIC TRANSIT SYSTEM SIMULATION

Deploying regenerative energy recuperation techniques in a given transit system must be preceded by a research step to identify and quantify the value propositions associated with such a deployment. Simulation is often used to compute the amount of energy consumption and peak power demand. Simulation results can assist engineers and decision makers to make an informed decision related to future investment, design and development. For instance, in the case of seeking optimal use of regenerative braking energy, simulation tools can help engineers determine the optimal technology, size, location, and other performance characteristics [81], [82].

Simulation of electric rail system has been studied since late 1970s [83]. There are various simulation studies and programs developed in both the academia and the industry. For instance, there are commercially available software packages, such as Vitas (a program developed for design and improvement of rail and signaling systems) [84], Trainops (a software package developed by LTK) [85] and Sitras Sidytrac by Siemens [86]. Another available program is Train Operation Model (TOM) [87] developed by Carnegie Melon University, which have three subroutines: (1) the Trains Performance Simulator (TPS); (2) the Electric Network Simulator (ENS); and (3) the Train Movement Simulator (TMS) [88]. Some other simulation tools developed by the academia include OpenTrack and Open Power Net by ETH university [89], [90], and Vehicle Simulation Program (VSP) by Vrije University of Brussel [91].

Generally, the simulation procedure of electrified transportation systems is based on load flow calculation. However, load flow calculation in this system is different from regular load flow analysis because of two main reasons [81], [92],[93]:

- The position of the electric load (i.e. the train) is changing over time
- Some parts of the system operate on DC power (i.e. the traction system) and some parts run on AC.

To perform load flow calculation, there are two main iterative techniques:

- The Gauss-Seidel method (easy implementation, but poor convergence)
- The Newton-Raphson method (fast convergence, but complex implementation)

Most of the work that has been done in load flow analysis of electrified transportation systems divide system modeling into two parts: (1) vehicle movement model, and (2) electric network model [94].

Simulation of train performance and that of traction power networks are dependent on each other. For example, an input for simulating traction power systems is train power demand, which is an output of train performance simulation. On the other hand, train motor performance is dependent on the traction voltage drop, and if the voltage drop level is significant, the amount of power demanded by a train will significantly decrease. Therefore, both train and network simulators must be coupled. In this case, train performance is simulated during a time period Δt and use the train state at time t and the voltage value from the previous time period, then the traction power network simulator uses the calculated power demand and calculates the voltage for the next time step [94].

There are two main categories for transient modeling of trains:

1) Cause-effect or forward facing method: In this method, the power consumed by the vehicle is used as an input to determine the speed of the wheel
2) Effect-cause or backward facing method: In this method, the speed profile and vehicle properties are used as inputs to determine the input power to the train.

To model electric rail vehicle with the "effect-cause" method, the speed of the train is taken as an input, and based on equations describing the vehicle dynamics the forces applied to the wheels are calculated. A schematic diagram of this approach is presented in Fig. 7.

Railway dynamics, equations of single train motion and load flow calculation of electric power supply, among other aspects, have been studied and presented in [81], [94], [95]–[100].

Most of the research work that has been carried out on modeling and simulating electrified transportation systems considers single train operation. Only a few studies considered multiple train operation. In [101], a Multi-train System Simulator (MTS) with discrete time update has been used as the main simulation core. Through this software, train movement and performance can be calculated and then, the Network Solver and the Network Capture Program are called. In the Power Capture Program, configuration of the power network and bus numbering are used as inputs to yield bus and line data for the Network Solver and the MTS. The Network Solver program's task is to perform power flow, bus voltage and power loss calculations. Another multi-train simulation tool that is called Simux has been presented in [83]. Through this simulation tool, regenerative braking can also be simulated. Another multi-train simulation method that is capable of modeling regenerative braking energy, and the various technologies that can be used to capture it has been presented in [102].

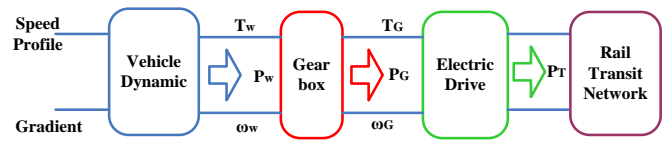

Fig. 7. Block diagram train performance modeling.

TABLE IX
COMPARISON OF VARIOUS SIMULATION METHODS

| Platform | Recovery Method | Comment | Ref. |
|---|---|---|---|
| MATLAB/SIMULINK | On board super Capacitor | Modeling single and multi-trams have been simulated. | [102] |
| - | Both onboard and wayside super capacitor | Modeling of two trains running on the same line has been presented. | [154] |
| MATLAB/SIMULINK | Wayside super capacitor | A control strategy for wayside super capacitor has been presented and simulated. | [155] |
| PSIM | Onboard super capacitor | A control strategy for super capacitors in electric vehicles has been proposed. | [156] |

Since the focus of this paper is on the recovery of regenerative braking energy, the simulation tools that are related to this subject have been summarized in Table IX.

## VIII. NONTECHNICAL ASPECTS

Even though the basic technical challenges are mostly solved, several nontechnical aspects can have a substantial influence on the integration of ESS into the existing railway systems. On the one hand, the systems need to be economically viable, which is often verifiable, but on the other hand, regulatory and energy pricing aspects may influence the level of transit system operator's engagement to deploy regenerative energy recuperation techniques. Some of these aspects will be discussed in this section, with focus on wayside ESS.

### A. Economic Aspects

Most rail transit cars that are built nowadays are already capable of producing regenerative energy during deceleration. However, the reuse of electricity may be limited. Actual reductions in energy use mainly depend on the number of start and stops as well as the traveled route [103]. To analyze the effectiveness of energy storage for capturing a larger share of the regenerative braking energy, many parameters need to be considered. The main aspect used for calculation of payback periods is often energy cost savings that can be accomplished by installing ESS. Additional revenue sources can be accomplished by participating in ancillary services. Payments to electric utilities for electric energy and peak power demand are, with up to 35%, a significant operating expense for Rail Transit Operators [69].

In this section, the potential economic benefit that can be gained by installing ESS in the electrical rail transit systems is briefly described and reviewed.

#### 1) Energy costs

Energy costs account for a considerable proportion of the operation costs of the mass transit systems and will become even more important as single urban rail trains become increasingly autonomous, so the labor cost will decrease substantially [104]. Because of the high dimension of traction energy, the energy price equals almost the wholesale price. To compare the potential for energy savings not just the pure traction energy costs, the costs for the transmission and distribution infrastructure has to be taken into account. For this reason, the average traction energy can be valued with approximately $110/MWh [2], [105].

#### 2) Conventional revenue streams

The grid based electrical energy demand reduction and regulating the voltage in the DC power grid are the basic revenue streams of ESS in railway application [106]. Moreover, another side advantage of deploying regenerative energy recuperation techniques is substation number reduction. According to [107], increasing the voltage at the train shoe enables an increased substation distance, thus reducing the number of substations. Although [107] was focused on TCR substations, the same effect can be achieved using reversible substations or wayside ESS.

TABLE X
APPLICATIONS FOR RAIL SYSTEM WAYSIDE ESS

| Application | Primary Objective | Type |
|---|---|---|
| Time-Based-Rate Management | Energy cost management | Economic |
| Demand Charge Management | Power cost management | Economic |
| Regenerative Braking | Energy Efficiency optimization | Economic |
| Renewable Energy Optimization | Resource optimization | Economic |
| VAR Control | Power quality improvement | Technical |
| Voltage Control | Power quality improvement | Technical |
| Peak Demand Reduction | Revenue optimization | Economic |
| Black start Capability | Revenue optimization | Economic |
| Renewable integration | Revenue optimization | Economic |

#### 3) Unconventional revenue streams

Today's unbundled electricity markets provide several additional possibilities to improve the profitability, mostly based on the power that can be provided to the electrical system. These opportunities can be found primarily in the area of electricity grid services, such as reactive power and voltage control, frequency control, operating reserves and black start capability. The following Table X shows some possible applications to increase the revenue of an ESS in rail transit networks apart from just reusing the regenerated energy from braking [106].

#### 4) Time based rate Management

If a transit agency has a time variant rate structure that varies through the day, month or year, the wayside ESS can be used to reduce the energy costs by charging them during off-peak rates and discharge during peak times. Certainly, this option is only viable if the difference between peak and off-peak times is big enough to compensate for the storage losses.

#### 5) Demand Charge Management – Peak Demand Reduction

The demand charge that a utility charges energy-extensive customers is usually billed once a month based on the maximal electric power use during a specified measurement interval (e.g. 15 minutes), with the intent to reduce the utility peak demand. A transit agency can reduce the cost of electricity and reduce the utility peak demand by supplying their energy, stored at lower use periods to the power gird. As in the previous case, the Demand Charge Management is just cost effective if the demand charge reduction compensates for energy losses. To get good and effective results, a combined optimization for time-based rate and demand charge management might offer an effective solution. Cost analyses pertinent to the use of battery, supercapacitor, and flywheel for peak demand reduction are presented in [108]–[111], respectively.

#### 6) Demand Response

As a distributed energy resource, the storage device can contribute to demand response programs corresponding to reliability issues or during times when marginal electrical energy prices are high, like typically on hot and humid summer days [112]. Hence, a change of the wayside ESS control can limit the infeed from the public power grid.

*7) Ancillary Services – Frequency Regulation*

Because of the common use case of wayside ESS, their ramp up time is often very short (i.e. seconds). For this reason, the installed systems are typically able to provide energy with low response times. To support the frequency regulation, wayside ESS can be used to bid in the wholesale frequency regulation market for "on call" regulatory services. This enables an additional revenue stream for wayside ESS by supporting the ISO in its role to organize the fine-tuning of the grid frequency. The exact use of wayside ESS for ancillary services varies depending on the regulations of the energy market.

*8) Emergency Power Supply*

Wayside ESS can also serve as a resource during emergency conditions. Commercial and industrial facilities are exposed to several issues like power quality problems or power outages, which could interrupt a production process or affect sensitive equipment that can cause production downtimes. Depending on the system and redundancy of a wayside ESS, providing emergency power supply can represent an additional revenue stream. Wayside ESS can be used to provide temporary backup power to allow safe shutdown of equipment in the event of a sustained or major outage, after a power interruption event. A backup power device can also allow for a successful transfer to a backup generator.

*9) Voltage and VAR Control*

To overcome power quality problems related to poor voltage control, wayside ESS can be used to stabilize voltage in the event of short-term fluctuations in the grid. Voltage sags can cause short-term problems and interruptions, but so repeated voltage excursions can lead to a reduced equipment life. The use of wayside ESS as voltage control units can reduce the replacement cost and provide savings through deferring substation upgrades . If the transport agency is operating its network on AC, the wayside ESS can be used as VAR control to overcome possible power quality issues. Since most of urban rail systems are currently operated with DC circuits, this mode is unlikely to provide a significant benefit for a typical transport agency.

*10) Renewable energy integration*

If a transit agency has intermittent renewable energy resources, the wayside ESS can be used to optimize the use of renewable resources. If time based rates apply, the wayside ESS can be a source to optimize the renewable portfolio.

*11) Smart Microgrid Integration*

The integration of a DC railway system into a DC micro grid solution might be another promising approach to increase the efficiency of urban rail systems. In [113], an approach was shown where the braking energy was stored and reused to charge public buses.

*B. Ownership Aspects*

A typical aspect in large infrastructure projects is the impact of various asset owners and policies. This means that the trains may just be operated by the transportation agency, but the ownership is by a third company, either an investment company or a company that produces or maintains transportation equipment. This can lead to different targets in terms of efficiency, because a licensed train operator may not always care about energy costs. On the other hand, there is a possibility that a third-party company can get licensed to build and operate a wayside ESS over a specific service time. Part of the revenue is diminished with this type of financing, but the first-time installation costs are not to be overtaken by the transportation authority [65].

*C. Reliability and Signaling*

To ensure a reliable train operation, it is essential to prove that the ESS and its signaling do not interfere with the signaling and communication systems of the transit system. For this reason, the potential interference of the ESS equipment in the frequency band of the signaling system should be studied. Because of the different technologies used in train signaling, the effects should be studied for each system to ensure the proper functionality. There is also a possibility of reversed interference. For this reason, the ESS needs to be secured against failures in the DC mains.

*D. Standardization*

Some general standards and guidelines exist for using energy storage in electric power systems; however, the electric rail transit system has specific features that may differ from the power utility system. Therefore, there is a need for a set of guidelines and standards specific for wayside energy storage technologies in electric rail transit systems.

Currently, there is a Draft Guide for Wayside Energy Storage for DC Traction Applications document, which covers the following topics: Normative references, Definitions, Applications, Common Technologies, Common Topologies, Specifying a Wayside Energy Storage System, Economic Considerations, Modeling and Simulation of Energy Storage, Performance, Safety and Environment, Installation and Integration and Verification and Validation [8].

## IX. CONCLUSIONS

In this paper, a comprehensive review on different methods and technologies the can be used for regenerative braking energy recovery has been presented. Three main solutions have been used worldwide including train timetable optimization, energy storage system, and reversible substations. Each application has its own pros and cons and can be implemented in different rail systems with different characteristics.

Train timetable optimization is a solution with low cost, which typically requires no new installations. The main purpose of this method is to synchronize the accelerating and decelerating phases of trains in order to increase the natural exchange of regenerative braking energy among them. This goal is achieved by optimizing the arrival, departure and dwell time of train operation. From the various studies reviewed in this paper, it can be seen that between 4% to a maximum of 34.5% energy saving has been claimed through timetable optimization. Besides maximizing utilization of regenerative braking energy, train timetable optimization can also reduce the peak power demand. However, application of this technique might be limited by certain service requirements. It requires supervisory real-time monitoring and control of the trains, which may not be available in some systems.

Another solution for regenerative energy recovery is through the use of energy storage systems. In this method, regenerative braking energy that is produced by trains is stored in onboard or wayside energy storage system, and released later on when it is needed. Beside energy savings, both types of ESS can reduce peak power demand and improve the third rail voltage level. Onboard ESS is mostly used when catenary free operation of the vehicle is needed; otherwise, due to the high cost of its implementation, wayside ESS is preferred. According to some publications, about 30% of the energy consumed by the train can be saved using ESS. The location, size and type of storage technology used for ESS significantly impact the amount of regenerative energy that can be recuperated. In addition, the various types of technologies that are available for storing regenerative braking energy have been reviewed in this paper. Super capacitors, batteries and flywheels are commonly used technologies. Among them, super capacitor has been used widely all over the world mostly because of its characteristics, such as fast response, high power density and long life cycle. In case of battery, Li-Ion battery is the most utilized one because of its high number of cycles, low weight, small size and commercial availability. A combination of these technologies seems to be a good option, but still needs further investigation.

Reversible substation is another way to increase the amount of energy that can be saved during vehicle braking. In this method, a reverse path is provided through an inverter for energy to flow back to the main grid. Implementing this method depends on the regulations of feeding power back to the main grid. There are two common methods to provide a revers path: a) combination of a diode rectifier with an inverter, b) using reversible thyristor-controlled rectifier (RTCR). There are commercially available reversible substations that are under test in several locations globally. When using reversible substation, priority should be given to the natural energy exchange between vehicles, and then, the surplus of energy can fed back to the main grid. Studies show that up to 13% of the consumed energy by a vehicle can feed back to the main grid using this method.

In order to make an optimal decision, before implementing any of the proposed solutions in an electric rail system, they have to be thoroughly analyzed. In this paper, some of the simulation tools and modeling techniques that have been developed by both the industry and academia, to perform this analysis, have been discussed.

Beside the technical aspect, some of the non-technical aspects pertaining to the deployment of regenerative braking energy recuperation techniques, with focus on the energy storage system solution were also discussed in this paper.

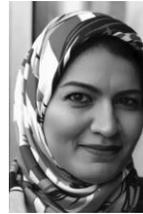

**Mahdiyeh Khodaparastan** received the B.Sc. and M.Sc degrees in electrical power engineering from Amirkabir University of Technology, Tehran, Iran, in 2010 and 2012, respectively. Currently, she is a PhD student and research assistant at the Smart Grid Laboratory, CUNY-City College, New York, USA. Her research interests include energy storage systems design and development, as well as power system protection and distributed generators.

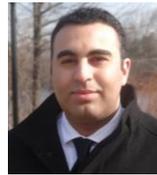

**Ahmed A. Mohamed (El-Tallawy)** (GS'09–M'13–SM'17) is an Assistant Professor of Electrical Engineering, and the Founding Director of the Smart Grid Laboratory, City University of New York, City College. He is also with the Department of Electrical Engineering (on leave), Menia University, Menia, Egypt. His current research interests include critical infrastructure interdependencies, smart grid resiliency, and electric transportation.

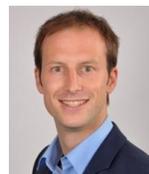

**Werner Brandauer** is a postdoctoral research fellow at the Smart Grid Laboratory, City University of New York, City College. He received his PhD degree in 2014 from Graz University of Technology. His main research interests include micro-grids, power system modeling, power system economics and renewable energy systems.